\documentclass[12pt]{article}                            
\usepackage{psfrag}
\usepackage{graphicx}                                                            
\usepackage{rotating}                                                            
\usepackage{longtable}                                                           
\usepackage{supertabular}                                                        
\usepackage{amsmath}

\begin{document}
\title{Importance of transverse dipoles in the stability of biaxial nematic phase: A Monte Carlo study \\[8mm]
} 
\author{Nababrata Ghoshal, Kisor Mukhopadhyay$^1$ and
Soumen Kumar Roy$^{2,}$
\footnote{Corresponding author. E-mail: skroy@phys.jdvu.ac.in,
Tel: +91 9874741525; fax: +91 33 24146584}\\
Department of Physics, Mahishadal Raj College,\\[1mm]
Mahishadal, Purba-Medinipur, West Bengal, India\\[1mm]
$^1$Department of Physics, Sundarban Mahavidyalaya,\\[1mm]
Kakdwip, South 24 Parganas, West Bengal, India\\[1mm]
$^2$Department of Physics, Jadavpur University,\\[1mm]
Kolkata - 700 032, India}
\date{  }
 
\maketitle
\begin{abstract}
 Monte Carlo simulation performed on a lattice system of
biaxial molecules possessing $D_{2h}$ symmetry and interacting 
with a second rank anisotropic dispersion
potential yields three distinct macroscopic phases depending on
the  biaxiality of the constituent molecules.
The phase diagram of such a system as a function of molecular biaxiality
is greatly modified when a transverse dipole is considered to be
associated with each molecule so that the symmetry is reduced 
to $C_{2v}$. Our results indicate the splitting of the
Landau point i.e. the point in the phase diagram where a direct
transition from the isotropic phase to the biaxial nematic phase occurs,
into a Landau line for a system of biaxial molecules with 
strong transverse dipoles.
The width of the Landau line becomes maximum for an optimal value of the 
relative dipolar strength.
The presence of transverse dipoles leads to the stabilization of 
the thermotropic biaxial nematic
phase at higher temperature and for a range of values of molecular biaxiality.
The structural properties in the uniaxial and biaxial phases are 
investigated by evaluating the first rank and second rank orientational 
correlation functions. The dipole induced long range order of 
the anti-ferroelectric structure in the biaxial nematic phase, is revealed.

\end{abstract}

\section{Introduction}
In the past few years great interest has been paid to thermotropic biaxial 
nematic liquid crystals, whose existence was first predicted by Freiser 
\cite{fr} in 1970 from mean field theory. Shortly afterwards, a number 
of theoretical \cite{al,st} and computer simulation \cite{luc,bis,camp}
 studies have been performed and phase diagrams consisting of three distinct 
macroscopic phases namely, the isotropic phase ($I$), the uniaxial nematic 
phase ($N_{U}$) (which may be, keeping the symmetry of the phase 
unchanged, of two types - prolate ($N_{U^+}$) or oblate ($N_{U^-}$)
nematic phases depending on the rod-like or disk-like nature of 
the molecules) and biaxial nematic phase ($N_{B}$) were generated. 
On the experimental front, scientists 
have been engaged in search for truly biaxial nematic phase 
in thermotropic liquid crystals since its theoretical prediction.
 Although there have been a number of reports \cite{mal,chan} of experimental 
identification of thermotropic $N_{B}$ phase since 1986,  none 
of these claims proved to be correct \cite{luck}. Recently there have been
claims of observing the elusive thermotropic $N_{B}$ phase for 
V-shaped molecules \cite{mad,ach} and for tetrapode molecules \cite{mer,fig}. 
These new findings have fuelled experimental and 
theoretical research \cite{bates1} on biaxiality 
in thermotropic liquid crystals.

\begin{figure}     
\begin{center}
\resizebox*{5cm}{!}{\includegraphics{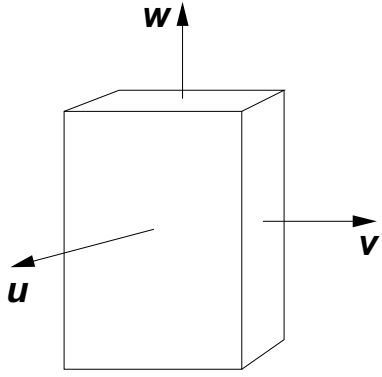}}
\caption{Schematic diagram of a biaxial molecule. Three molecular symmetry axes  are $\boldsymbol{u}$, $\boldsymbol{v}$ and $\boldsymbol{w}$.}\label{f1}
\end{center}
\end{figure}

The molecules forming liquid crystals always deviate from their 
assumed cylindrical symmetry and their usual structure can 
schematically be described as board-like (see Figure $\ref{f1}$).
Usually this type of molecules give rise to the $N_{U}$ phase as a 
consequence of the long range orientational order of the molecular 
long symmetry axis ($\boldsymbol{w}$) while the short symmetry axes 
($\boldsymbol{u}$, $\boldsymbol{v}$) remain uncorrelated except at short range.
In this case the molecular long axes tend to be parallel, on average, 
along a single macroscopic direction called the principal director 
($\boldsymbol{n}$). At some state point long range orientational ordering 
is established for the molecular long axes as well as for the molecular 
short axes which results in a biaxial nematic phase and the corresponding 
average preferential alignments of the short axes occur along the two 
secondary directors ($\boldsymbol{l}$, $\boldsymbol{m}$).

The ability of a material to form the $N_{B}$ phase depends on the 
molecular biaxiality of the constituent molecules. 
In the case of isotropically averaged dispersive interactions 
(which we have used in the present work) there is a single 
molecular biaxiality parameter $\lambda$ \cite{longa, rbera} 
which is  related to the anisotropy of the polarizability 
tensor of the molecule. 
From the study of molecular field theory \cite{luck} it has 
been observed that for obtaining 
a stable $N_{B}$ phase, $\lambda$ must have values within 
a very narrow range around the optimal biaxiality ($\lambda_C$).
 The theoretical value of $\lambda_C$ is $1/\sqrt{6}$ ($\approx 0.40825$),
 which corresponds to the transition point from a system of 
prolate to oblate molecules \cite{bis,luck,chic}. 
For this value a second-order transition occurs directly 
between biaxial and isotropic phases and the corresponding point in the 
phase diagram is known as the Landau point. At this special critical point 
where two second order critical lines meet a first order phase boundary, 
$N_B$ phases are expected to appear at the highest temperature and 
therefore, on the basis of theoretical analysis one may predict that 
to be a good candidate towards forming the $N_{B}$ phase the molecule 
should have the biaxiality very close to $\lambda_C$. But it is almost 
an impossible task for the experimentalists to design mesogenic molecules 
with the desired biaxiality due to molecular flexibility.
However, influence of other factors, such as polydispersity \cite{pol}, large 
transverse electric dipoles associated with the mesogenic molecules 
etc., enhances the stability of the $N_B$ phase.  
The present scenario in context of the transverse dipole moments is 
elaborated below. 

For symmetric bent-core molecules (V-shaped) the 
optimal biaxiality factor $\lambda_C = 1/\sqrt{6}$ occurs for a single 
angle of $109.47^\circ$ between the arms which is the tetrahedral angle.
On the other hand Madsen et.~al. \cite{mad} while investigating  a class of 
bent-core molecules with different angles between the arms claimed to 
have observed a biaxial phase at an inter-arm angle as large as $140^\circ$.
At this angle the molecular biaxiality factor is such that a $N_U$ phase 
is expected. Although the work of Madsen et.~al. \cite{mad}  
does not give a conclusive evidence of the occurrence of a biaxial nematic, 
it may be noted that the type of molecules they considered 
had a strong transverse electric dipole moment. 
To find if this could in some way make the job of finding 
the biaxial nematic phase easier, Bates \cite{bates2} performed 
a Monte Carlo simulation using a lattice model of bent-core 
molecules with transverse dipoles attached to the centre. In this 
work each molecule was assumed to have two symmetric arms and the 
included angle 
was varied. Each arm of a molecule interacts with each of the twelve 
arms of the six nearest neighbour molecules confined to a simple cubic 
lattice via the usual Lebwohl-Lasher potential \cite{lebo}. 
In addition, the dipole-dipole interaction, which actually is of long range 
nature, was also confined to the nearest neighbours for simplicity and 
was taken to be of the $-\cos\beta$ type where $\beta$ is the angle between 
the interacting dipoles. The phase diagram that Bates obtained was quite 
different from the usual one obtained while treating non-polar bend-core 
molecules where one gets a single Landau point for a bend angle of 
$109.47^\circ$. Instead, he observed that for large dipole moments 
the single Landau point gets transformed into a series of Landau points
usually called a Landau line which covers a significant range of angles
(and hence a range of molecular biaxiality) across which a direct 
$N_{B}$ $\leftrightarrow$ $I$ first order transition occurs. 
If this is really the case 
then the job of the experimentalists becomes a little easier.

On the theoretical front there is a very recent work on bent-core 
molecules by Grzybowski and Longa \cite{grzy}
who used density functional theory (DFT) to analyze 
two- and three-segment Gay-Berne (GB) molecules
with a single transverse dipole moment. 
They found that nonpolar models with two uniaxial arms has 
a single Landau point 
and the inclusion of the dipoles shifts the Landau point towards lower apex 
angles for the system containing two-arm molecules. For the three-segment
molecules with large dipole moments a Landau line occurs and they observe
that for an optimal dipole strength the $N_B$ phase is most probable.
In all cases these authors show that further strengthening of the 
dipolar interaction results in shrinking the distance between the 
isotropic phase and the biaxial nematic phase which indicates that the 
intermediate uniaxial nematic phase becomes less stable.
The study reveals that in the most likely situation for detecting the $N_B$
phase each molecule should consist of three segments with a transverse 
dipole moment of about $3D$.

In this communication we present the results of a Monte Carlo simulation
in a lattice model where biaxial molecules of $D_{2h}$ symmetry interact 
with nearest neighbours via an anisotropic dispersion potential. 
To study the effect of dipolar interaction on the phase behaviour of 
the system we have associated a transverse dipole moment with each molecule
which reduces the symmetry of the molecules to $C_{2v}$. 
The dispersion interaction has been extensively used in mean field and 
MC studies \cite{luc,bis,chic} to predict the phase behaviour of nematic liquid crystals.
The case of symmetric V-shaped molecules considered in the 
work of Bates \cite{bates2}
can be deduced from the dispersion potential we have used, as has been
demonstrated in the work of Romano \cite{roma2} and 
Bates and Luckhurst \cite{bates1}.
Although we have used the isotropically averaged out (over the 
intermolecular vector) form of the dispersion potential (which is short 
range in nature), the dipolar interactions have been considered in the 
full form and keeping in mind the long range nature of this interaction 
we have used the reaction field (RF) method \cite{bark, tild} in 
our calculation to improve the reliability of the results.
Although the RF method is less rigorous than the Ewald summation (ES)
method, it allows much computational advantages than the latter and
the difference of the averages of thermodynamic quantities
obtained by the RF and ES techniques are within the statistical
uncertainty of the simulation \cite{gil, hou, bera}.
 We have compared our findings with the 
results of Bates \cite{bates2} and Grzybowski and Longa \cite{grzy} 
and these seem to be in good qualitative agreement. 
It may be noted that the dispersion potential we have used is a 
little more general than that in \cite{bates2} since
the former also includes the case of V-shaped molecules with non-symmetric arms.
We have also investigated the details of the phase structures of this model
by evaluating structural quantities like first-rank and second-rank 
correlation functions.
  
\section{The Model}
 In our study we consider identical biaxial molecules with a transverse 
dipole, whose centers of mass are associated with a simple-cubic lattice. 
The interaction energy of two molecules $i$ and $j$ is given by 
\begin{equation}\label{e1} 
U_{ij}=U_{ij}^{disp}+U_{ij}^{\mu\mu}
\end{equation}
where $U_{ij}^{disp}$ is the pair potential obtained from London
dispersion model \cite{buck,stone} and $U_{ij}^{\mu\mu}$ is the dipole-dipole 
interaction. The dispersion term decays much faster   
(as $1/r^6$) than the dipolar term (which varies as $1/r^3$). We 
therefore, restrict 
the dispersion interaction to the six nearest neighbours.  
The orientationally anisotropic dispersion interaction explicitly 
depends both on the mutual orientation of the two interacting molecules, 
 and on their orientations with respect to the  
intermolecular unit vector ($\boldsymbol{\hat{r}}_{ij}$).    
By isotropically averaging over the intermolecular 
unit vector, $\boldsymbol{\hat{r}}_{ij}$ the dispersion potential between two 
identical neighbouring molecules becomes 
\begin{equation}\label{e2} 
U_{ij}^{disp}=-\epsilon_{ij} \{R_{00}^2(\Omega_{ij})+2\lambda[R_{02}^2(\Omega_{ij})+R_{20}^2(\Omega_{ij})]+4\lambda^2R_{22}^2(\Omega_{ij})\}.
\end{equation}
Here $\Omega_{ij}=\{\phi_{ij},\theta_{ij},\psi_{ij}\}$ denotes the triplet 
of Euler angles defining the relative orientation of $i^{th}$ 
and $j^{th}$ molecules; we have used the convention used by Rose \cite{rose}
 in defining the Euler angles. 
$\epsilon_{ij}$ is the strength parameter which is assumed 
to be a positive constant ($\epsilon$) when the particles $i$ and $j$ are 
nearest neighbours and zero otherwise. $R_{mn}^L$ are combinations of 
symmetry-adapted ($D_{2h}$) Wigner functions 
\begin{equation}\label{e3}
R_{00}^2=\frac{3}{2}\cos^2\theta-\frac{1}{2}
\end{equation}  
\begin{equation}\label{e4}
R_{02}^2=\frac{\sqrt{6}}{4}\sin^2\theta\cos2\psi
\end{equation}  
\begin{equation}\label{e5}
R_{20}^2=\frac{\sqrt{6}}{4}\sin^2\theta\cos2\phi
\end{equation}  
\begin{equation}\label{e6}
R_{22}^2=\frac{1}{4}(1+\cos^2\theta)\cos2\phi\cos2\psi-\frac{1}{2}\cos\theta\sin2\phi\sin2\psi.
\end{equation}  

The parameter $\lambda$ is a measure of the molecular biaxiality and depends
on the molecular properties. For the dispersion interactions, $\lambda$ can be 
expressed in terms of the eigenvalues ($\rho_1$, $\rho_2$, $\rho_3$) of the 
polarizability tensor {\boldmath $\rho$} of the biaxial molecule
\begin{equation}\label{e7}
\lambda=\sqrt{\frac{3}{2}}\frac{\rho_2-\rho_1}{2\rho_3-\rho_2-\rho_1}.
\end{equation}
The condition for the maximum biaxiality is 
$\rho_3-\rho_2=\rho_2-\rho_1>0$ and $\lambda=\lambda_C=1/\sqrt{6}$ 
(this self-dual geometry corresponds to the Landau point in the 
phase diagram). $\lambda < \lambda_C$ corresponds 
to the case of prolate molecules whereas $\lambda > \lambda_C$ corresponds
to oblate molecules.
The V-shaped lattice model used by Bates \cite{bates2} is equivalent 
to the present dispersion model and it can be shown \cite{bates1} that 
both the models produce identical results when the temperature 
is rescaled properly by using
\begin{equation}\label{e8}
T^\ast=\frac{4T_{VSM}^\ast}{(1-3\cos\gamma)^2}
\end{equation}
and the molecular biaxiality parameter is given by   
\begin{equation}\label{e9}
\lambda=\sqrt{\frac{3}{2}}\frac{1+\cos\gamma}{1-3\cos\gamma}
\end{equation}
where $\gamma$ represents the angle between the two symmetric arms 
of the V-shaped molecule. $T^\ast$ and $T_{VSM}^\ast$ are respectively the 
dimensionless temperatures used in this model and in the V-shaped 
symmetric model studied by Bates. 
In general, the biaxiality parameter 
$\lambda$ depends both on the inter-arm angle and the relative mesogenic
anisotropy of the arms for the symmetric as well as 
non-symmetric V-shaped molecules \cite{bates1}.

The fully anisotropic biaxial dispersion model  
has been studied for both two-dimensional
and three-dimensional lattices by mean-field and MC methods \cite{roma1, roma2}.
The MC simulations performed on a three-dimensional simple cubic lattice 
system of biaxial molecules interacting via the fully anisotropic 
dispersion model in one case and its isotropically averaged form 
in the other case suggest that both the models produce almost similar 
results \cite{roma2}. We, therefore, opted the simpler form of the 
biaxial dispersion pair potential (Eq. ($\ref{e2}$)) which has been
investigated in previous studies and phase diagrams for a fairly wide range
of molecular biaxiality have been generated \cite{bis,chic}.  

The dispersion potential (Eq. ($\ref{e2}$)) can also be expressed in 
Cartesian form \cite{chic} as 
\begin{equation}\label{e10}
U_{ij}^{disp}=-\epsilon \{[\frac{3}{2}(\boldsymbol{w}_i\cdot\boldsymbol{w}_j)^2-\frac{1}{2}]-\lambda\sqrt{6}[(\boldsymbol{u}_i\cdot\boldsymbol{u}_j)^2-(\boldsymbol{v}_i\cdot\boldsymbol{v}_j)^2]+
\lambda^2[(\boldsymbol{u}_i\cdot\boldsymbol{u}_j)^2+(\boldsymbol{v}_i\cdot\boldsymbol{v}_j)^2-(\boldsymbol{u}_i\cdot\boldsymbol{v}_j)^2-(\boldsymbol{v}_i\cdot\boldsymbol{u}_j)^2]\}
\end{equation}
where $\boldsymbol{u}$, $\boldsymbol{v}$ and $\boldsymbol{w}$ are the three mutually orthogonal 
unit vectors along the molecular symmetry axes with the convention \cite{rosso} that 
$\boldsymbol{u}$ and $\boldsymbol{w}$ are along the shortest and longest axes 
respectively (Figure $\ref{f1}$).
Here the orientation of each biaxial molecule, in terms of the Euler angles, 
 with respect to the laboratory 
frame $\{\boldsymbol{e_x},\boldsymbol{e_y},\boldsymbol{e_z}\}$ is given by, 
\[
\boldsymbol{u}=(\cos\theta\cos\phi\cos\psi-\sin\phi\sin\psi)\boldsymbol{e_x}\\
+(\cos\theta\sin\phi\cos\psi+\cos\phi\sin\psi)\boldsymbol{e_y}-
\sin\theta\cos\psi\boldsymbol{e_z},
\]
\[
\boldsymbol{v}=-(\cos\theta\cos\phi\sin\psi+\sin\phi\cos\psi)\boldsymbol{e_x}\\
-(\cos\theta\sin\phi\sin\psi-\cos\phi\cos\psi)\boldsymbol{e_y}+
\sin\theta\sin\psi\boldsymbol{e_z},
\]
and
\[
\boldsymbol{w}=\sin\theta\cos\phi \boldsymbol{e_x}+\sin\theta\sin\phi 
\boldsymbol{e_y}+\cos\theta \boldsymbol{e_z}.
\]
The dipole-dipole interaction for a pair of point dipoles is given by
\begin{equation}\label{e11}
U_{ij}^{dd}=  \frac{\mu^2}{4\pi\epsilon_0\sigma^3r_{ij}^3}[(\boldsymbol{u}_i\cdot \boldsymbol{u}_j)-3(\boldsymbol{u}_i\cdot \boldsymbol{\hat{r}}_{ij})(\boldsymbol{u}_j\cdot \boldsymbol{\hat{r}}_{ij})]. 
\end{equation}
Here $\boldsymbol{\hat{r}}_{ij}$ is a unit vector along the 
inter-molecular vector 
connecting the centres of mass of molecules $i$ and $j$ of the dimensionless 
separation $r_{ij}$,
$\mu$ is the common value of the transverse dipole moment, $\sigma$ is 
the nearest neighbour separation and 
$\boldsymbol{u}_i$ is the unit 
vector along the shortest molecular symmetry axis of the $i^{th}$ molecule. 
The long-range dipolar potential contributions to the total pair 
potential have been evaluated using the reaction field (RF) method.
In this method each dipole is assumed to interact with all others within 
a spherical cavity of radius $r_c$ and beyond this cutoff radius the 
dipoles are considered to act as a dielectric continuum of dielectric 
constant $\epsilon_s$ producing a reaction field within the cavity.
Therefore the dipolar pair potential within RF geometry can be 
written as
 
\begin{equation}\label{e12}
U_{ij}^{\mu\mu}=  U_{ij}^{dd}-\frac{2(\epsilon_s-1)\mu^2}{2\epsilon_s+1}\frac{(\boldsymbol{u}_i\cdot \boldsymbol{u}_j)}{4\pi\epsilon_0\sigma^3r_c^3}, 
\end{equation}
for $r_{ij}<r_c$ and $U_{ij}^{\mu\mu}=0$ for $r_{ij}>r_c$.
The truncation must be done for $r_c<L/2$, $L$ being the lattice dimension
used in the simulation.
The dielectric constant $\epsilon_s$ of the surrounding medium is usually 
chosen to be one of the two extremes, $1$ for vacuum or $\infty$ for 
conductor. But a self consistent method \cite{gil} for estimating 
$\epsilon_s$ during the simulations is most appropriate in the RF method. 
We have used a parameter 
\begin{equation}\label{e13}
\epsilon^\ast=\frac{\mu^2}{4\pi\epsilon_0\epsilon\sigma^3}
\end{equation}
 which determines the relative strength 
of the dipolar interaction to the dispersion interaction. For a given value 
of $\epsilon^\ast$ the relative contribution of the dipolar energy in the 
total energy at the lowest temperature can be estimated by evaluating 
the ratio $2.676\epsilon^\ast/(4.0+2.676\epsilon^\ast)$ where the 
numerator is related to the negative of the ground-state energy 
(per particle) of a simple cubic dipolar lattice system \cite{roma3} 
whereas the first term in the denominator is the negative of the 
lowest value (per particle) of the dispersion energy for the optimal 
value of the molecular biaxiality $\lambda=1/\sqrt{6}$. 
The dimensionless temperature used is defined as $T^\ast=k_BT/\epsilon$.

\section{Computational Aspects}
\label{compu}
To explore phase diagrams we performed a series of Monte Carlo (MC) simulations 
using the conventional Metropolis algorithm
on a periodically repeated simple cubic lattice model, consisting of 
$N=40^3$ particles. Simulations were run in cascade, in order of increasing 
dimensionless temperature $T^\ast$. Equilibration runs were typically of 
$20\,000$ sweeps where each sweep consisted of $N$ MC steps (or moves)
and were followed by a production run of $20\,000 - 30\,000$ sweeps 
(longer runs were used close to the transitions).

In our case a Monte Carlo move was attempted by selecting a site at random and 
then by choosing one of the laboratory axes at random the  
molecule at that site was rotated about the chosen laboratory axis following 
the Barker-Watts method \cite{bar}.   
The dipolar energy of the molecules has been computed using the RF method, 
with cut off $r_c=8$ (in units of lattice spacing). 
The effect of the cut off radius on the simulation results have been studied
in \cite{morro, stein}. In our case further increase in $r_c$  produces almost 
indistinguishable results but increases the computation time significantly. 
For example if we choose $r_c=10$ the averaged quantities  
like order parameters and internal energy change by less than $0.1\%$   
compared to the results obtained for $r_c=8$.
However, the CPU time was $62$ hours for the former case 
and is about $28$ hours for $r_c=8$ for simulation at a given 
temperature with $20\,000$ Monte Carlo steps per site on 
Intel Core $i7$ $960$ processors clocked at $3.2$ GHz. 
The dielectric constant $\epsilon_s$ 
has been evaluated using a self consistent method \cite{gil}.
In order to analyze the orientational order we have calculated the order
parameters $\langle R_{mn}^2\rangle$ following the procedure described
in \cite{allen}. According to this, a $\boldsymbol{Q}$ 
tensor is defined for the 
molecular axes associated with a reference molecule. For an arbitrary unit 
vector {\boldmath $w$}, the elements of the $\boldsymbol{Q}$ 
tensor are defined as 
\begin{equation}\label{e14}
Q_{\alpha\beta}(\boldsymbol{w}) = \langle(3 w_\alpha w_\beta-\delta_{\alpha\beta})/2\rangle
\end{equation}
where the average is taken over configurations and the subscripts $\alpha$ 
and $\beta$ label Cartesian components of {\boldmath $w$} w.r.t the 
laboratory frame. The three $\boldsymbol{Q}$ tensors associated 
with three molecular symmetry axes are diagonalized 
(once after each MC sweep) and the eigenvalues and 
the eigenvectors obtained are then recombined to give the order parameters 
$\langle R_{mn}^2\rangle$ w.r.t the director frame \cite{allen}. 

Of the four second rank order parameters only $\langle R_{00}^2\rangle$
 and $\langle R_{22}^2\rangle$ are of importance 
in practice. $\langle R_{00}^2\rangle$ is the usual uniaxial order parameter
$\langle P_2\rangle$ which measures the alignment of the longest 
molecular symmetry axis with the primary director ($\boldsymbol{n}$) and $\langle R_{22}^2\rangle$ is the measure of 
the alignment of the transverse molecular symmetry axes 
 about the secondary directors ($\boldsymbol{l}$,
$\boldsymbol{m}$). $\langle R_{00}^2\rangle$ is zero in an 
$I$ phase and it is non-zero in the $N_{\it U}$ phase as well as 
in the $N_{\it B}$ phase with maximum value $1$
in the perfectly ordered state. On the other hand 
$\langle R_{22}^2\rangle$ is zero in the $I$ phase
and also in the $N_{\it U}$ phase and is non-zero only 
in the $N_{\it B}$ phase and increases towards its maximum value 
of $0.5$ in perfectly ordered biaxial phase.  

 We have also calculated the reduced specific heat per particle
from fluctuations in the energy
\begin{equation}\label{e15}
C_V^\ast = \frac{\langle {E}^2 \rangle-{\langle E \rangle}^2}{Nk_B{T}^2}
\end{equation}
where $E$ is the total energy of the system.

Further details on the phase structure of the  $N_{\it B}$ phase
in presence of transverse dipoles can be found by analyzing structural 
quantities, like the first-rank orientational correlation function 
defined by 
\begin{equation}\label{e16}
g_1(r) = \langle P_1(\boldsymbol{u_1}\cdot \boldsymbol{u_2})\rangle_r,
\end{equation}
and the second-rank orientational correlation function,
\begin{equation}\label{e17}
g_2(r) = \langle P_2(\boldsymbol{u_1}\cdot \boldsymbol{u_2})\rangle_r.
\end{equation}
The first, which is essentially just the average cosine of the angle between 
the transverse symmetry axes of 
molecule $1$ and  molecule $2$ separated by a distance $r$,
is used to determine the anti-ferroelectric structure of the biaxial phase, 
while $g_2(r)$ is useful in assessing the ordering of the transverse 
molecular symmetry axes irrespective of their parallel or antiparallel 
arrangements.

\section{Results and discussion}
We present phase diagrams for five different values of the relative 
strength factor $\epsilon^\ast=0, 0.4, 0.8, 1.2$, and $1.6$. 
The different phases are identified from the non-vanishing values of 
the relevant order parameters while the transition temperatures 
are obtained from the positions of the peaks in the temperature 
dependence of the specific heat curves ($C_V^\ast$) as well as 
from the order parameter curves and these were found to match
with each other within the errors present in the simulation.
In the neighbourhood of a transition we have taken numerical
data at temperature intervals of 0.025 and elsewhere the intervals
were greater. To make a conclusion about the order of the phase 
transitions we have used the order parameter curves -- for  
continuous vanishing of an order parameter the transition was taken 
as second order whereas a jump to zero or near zero value (which is a 
finite size effect) from a finite value was considered to be first order 
transition. These observations were in general supported by our 
results on specific heat plots. At a second order phase transition 
the peak heights are relatively smaller than those in a first order 
transition where the peaks are narrow and sharp. It may be noted 
that this method of identifying the phase transition temperatures 
and the order of the transition is only qualitatively correct.
To ascertain these accurately one needs to perform more extensive 
MC simulation using, for instance, the finite size scaling methods
which involve collecting data on systems of different size. We did 
not go into this exercise and therefore expect our results only to be 
qualitatively correct.

\begin{figure}
\begin{center}
\resizebox*{100mm}{!}{\includegraphics{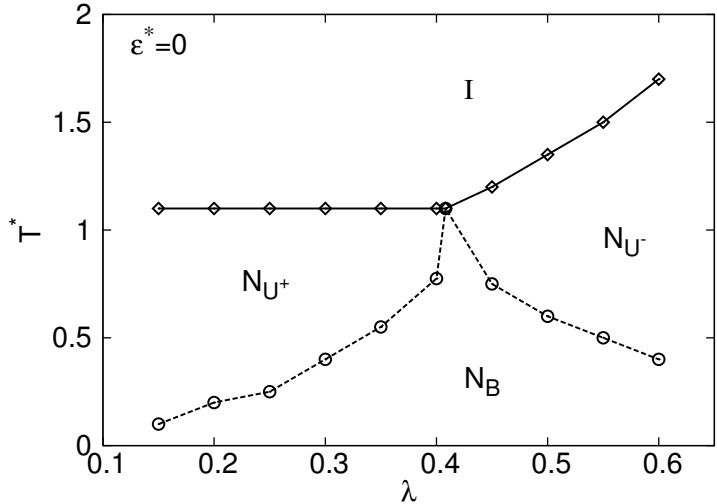}}
\caption{The phase diagram as a function of dimensionless temperature $T^\ast$
and molecular biaxiality $\lambda$ (ranging from $0.15$ to $0.6$ in 
step of $0.05$ along with the self-dual value of $0.40825$) 
for a system of biaxial molecules
interacting via  second rank anisotropic dispersion potential only. The
solid line represents the first-order transitions, while the dashed line
represents the second-order transitions.}\label{f2}
\end{center}
\end{figure}

\begin{figure}
\begin{center}
\resizebox*{100mm}{!}{\includegraphics{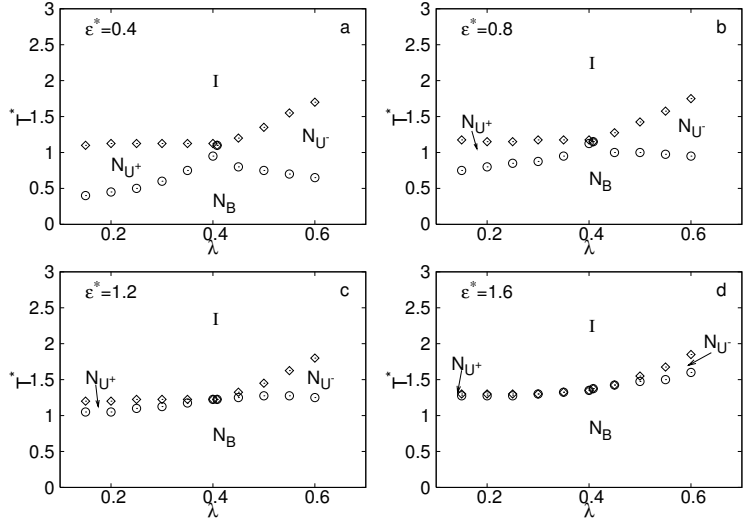}}
\caption{The phase diagrams of biaxial molecules interacting via  the
resultant potential with four different dipolar strengths $\epsilon^\ast=
0.4,0.8,1.2,1.6$.}\label{f3}
\end{center}
\end{figure}

In Figure \ref{f2} we have shown the phase diagram for $\epsilon^\ast=0$
which means that only the isotropically averaged dispersion interaction 
(confined to nearest neighbours) given by Eq. (\ref{e2}) has been 
considered. The result is same as obtained by Biscarini et al
\cite{bis} and shows the $I$, 
$N_{U^+}$, $N_{U^-}$  and the $N_B$ phases.
The biaxiality factor $\lambda$ has been varied from $0.15$ to $0.6$.
The $N_{B}$ $\leftrightarrow$ $N_{U}$ transition is found to be 
second-order while $N_{U}$ $\leftrightarrow$ $I$ transition is first-order
in nature and these are well known from previous 
studies \cite{al,st,bis,luck}.
At the Landau point (i.e. for $\lambda =1/\sqrt{6}\approx 0.40825$) a direct 
$N_{B}$ $\leftrightarrow$ $I$ transition takes place and this is 
second-order in nature.
In Figure \ref{f3} we have shown the phase diagrams when the long range 
interactions among the transverse dipoles have been considered in addition 
to the dispersion potential. The diagrams have been plotted for four 
values of the relative dipolar strength $\epsilon^\ast=0.4, 0.8, 1.2,$ 
and $ 1.6$. When the strength of the dipolar interaction is weak 
($\epsilon^\ast=0.4$ and  $0.8$) 
the higher transition 
temperature $T^\ast_{N_UI}$ remains almost unaltered but the presence of
(transverse) dipolar interaction stabilizes the $N_B$ phase by raising 
the temperature $T^\ast_{N_BN_U}$ to higher value. By increasing 
the relative dipolar strength, $\epsilon^\ast$ to $1.2$ 
an indication of the formation of Landau line is observed and for 
strong dipolar interaction ($\epsilon^\ast=1.6$) 
the $N_{B}$ $\leftrightarrow$ $I$
transition occurs for a range of values of the biaxiality factor $\lambda$
(from $0.3$ to $0.45$) instead of a single value.
Thus the Landau point (for $\epsilon^\ast=0$) 
turns into a Landau line in presence of strong transverse dipole moments 
(Figure \ref{f3}d). Besides performing detail study for $\epsilon^\ast$ 
upto $1.6$ we have performed a search for the extension of the Landau
line for $\epsilon^\ast=2.0, 2.4$ and $2.8$, where simulations were 
performed only to locate the two ends of the Landau lines.
We have observed, that the formation of Landau 
line starts at $\epsilon^\ast=1.2$ and the width of the Landau 
line i.e. range of the values of $\lambda$ showing 
the direct $N_{B}$ $\leftrightarrow$ $I$ transition, first increases 
and becomes maximum extending from $\lambda = 0.15$ 
to $0.5$ for $\epsilon^\ast=2.0$ at which the dipolar interaction
is approximately $57\%$ of the total energy at very low temperature. 
For further increase in the dipolar strength 
the width of the direct $N_{B}$ $\leftrightarrow$ $I$ transition becomes 
shorter and gets shifted towards higher values of $\lambda$. For 
example when  $\epsilon^\ast=2.4$, the Landau region covers the range of 
biaxiality parameter from $\lambda = 0.35$ to $0.55$ and 
for $\epsilon^\ast=2.8$ it extends from $\lambda = 0.55$ to $0.7$.

We make a qualitative comparison of our findings with those 
of  Grzybowski and Longa \cite{grzy}. 
These authors used GB interaction between two-segment and 
three-segment molecules and also considered the effect of attaching 
a transverse dipole to each molecule. For two-segment molecules they do
not find any Landau line resulting from the presence of the dipoles.
This is in disagreement with our observation and 
with those of Bates \cite{bates2}.
Moreover, we observe that after the formation of Landau line,
with increase in the strength of the dipoles the extension of the 
Landau line first increases, reaches a maximum and then shrinks,
always moving towards higher values of $\lambda$. 
Grzybowski and Longa \cite{grzy} for three-segment GB molecules 
find the same features in the extension of the 
Landau line. But initially this happens for lower values of the apex 
angle $\gamma$ (i.e. higher $\lambda$) and beyond a certain value of 
the dipole strength the Landau line, which is of smaller extension, 
occurs at higher values of $\gamma$. 
Moreover, we observe that the most
favourable situation (in terms of the width of the Landau line) 
for the occurrence of the $N_B$ phase occurs for $\epsilon^\ast=2.0$.
If one considers the lattice spacings ($\sigma$) to be of the order of 
$5 \AA$ and uses the relation in Eq. (\ref{e13})
\cite{roma4}, then for a transition temperature of $473$ K 
(which is close to that of the V-shaped molecule used in \cite{mad}),  
the estimated value of the dipole moment turns out to be $3.3$ D 
which is close to the value of $\mu$ obtained in \cite{grzy}. 
Here we have used the reduced transition temperature to be $1.5$
to estimate $\epsilon$.

We now make a comparison of our results with those of Bates \cite{bates2}.
Bates observed that as the dipolar strength is increased, there is an 
increase in the $N_{B}$ $\leftrightarrow$ $N_U$ transition temperature.
For $\epsilon^\ast=0.5$ the temperature at which the $N_B$ phase is 
observed is essentially independent of $\gamma$ (or, $\lambda$) and 
this is just below the temperature at which the Landau point is observed
for $\epsilon^\ast=0$. This feature is similar to our phase diagram 
shown in Figure \ref{f3}c for $\epsilon^\ast=1.2$. We are also inclined 
to state that for this value of the dipolar strength we find that the 
Landau line just begins to appear. Moreover, Bates' work shows that 
for $\epsilon^\ast=1.0$ there is a Landau line extending 
from $\gamma = 107^\circ$ to $122^\circ$ (the corresponding 
biaxiality parameter $\lambda$ being from $0.46$ and $0.22$) across 
which there is a first order $N_{B}$ $\leftrightarrow$ $I$ transition.
This may be compared with our observation for $\epsilon^\ast=1.6$ where
we find a Landau line extending from $\lambda=0.30$ to $0.45$ across
which there is a first order $N_{B}$ $\leftrightarrow$ $I$ transition.
We however point out that a direct comparison of our results 
with those of Bates is not meaningful because of different features 
appearing in the phase diagrams for different values of the 
dipolar strengths. This is not surprising since the treatment 
of the long range dipolar interactions using the RF method rather 
than truncating it at the nearest neighbours is likely to make 
this difference. To state this more clearly, the qualitative features 
observed by Bates now appear at an enhanced value of $\epsilon^\ast$.       
Also, while comparing the two sets of results, we need to rescale 
the temperatures using Eq. (\ref{e8}). 
   
\begin{figure}
\begin{center}
\resizebox*{100mm}{!}{\includegraphics{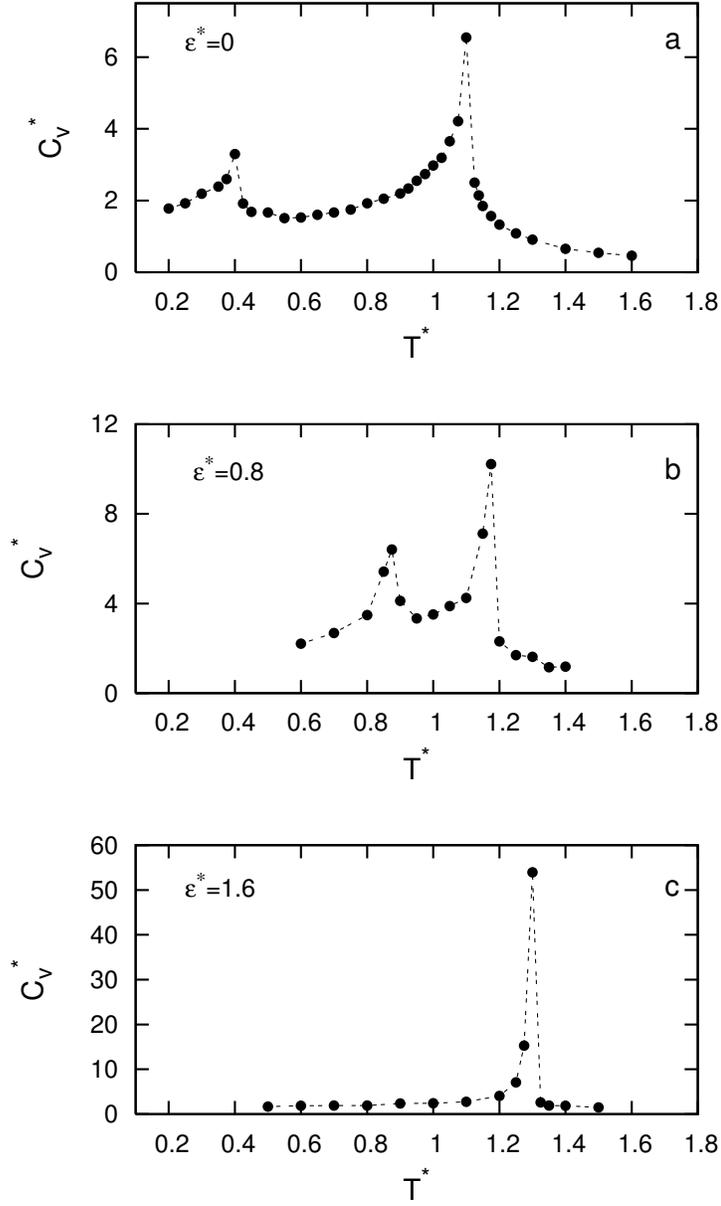}}
\caption{Specific heat per particle vs dimensionless temperature $T^\ast$
 for the molecular biaxiality $\lambda=0.3$ and for three different
 dipolar strengths $\epsilon^\ast= 0,0.8,1.6$. }\label{f4}
\end{center}
\end{figure}

\begin{figure}
\begin{center}
\resizebox*{100mm}{!}{\includegraphics{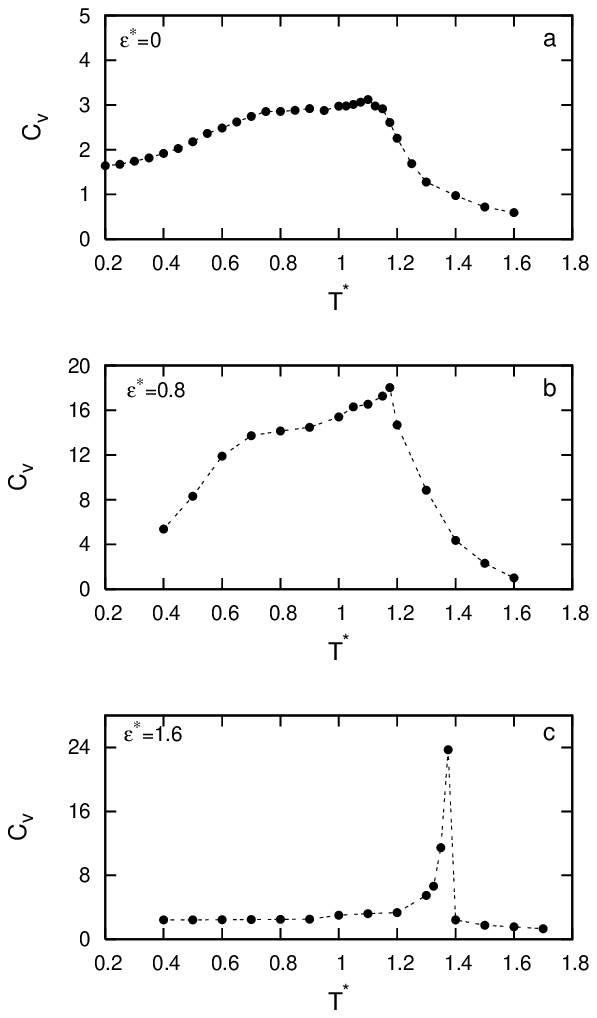}}
\caption{Specific heat per particle vs dimensionless temperature
$T^\ast$ for the molecular biaxiality parameter $\lambda=0.40825$ and
for three different dipolar strengths $\epsilon^\ast= 0,0.8,1.6$.}\label{f5}
\end{center}
\end{figure}

In Figures \ref{f4} and \ref{f5} we show the plots of
$C_V^\ast$ for $\lambda = 0.3$, and $\lambda = 0.40825$ respectively
for the case of pure dispersion interaction ($\epsilon^\ast=0$)
and for two other cases
with relative dipolar strengths $\epsilon^\ast=0.8$ and $1.6$.
In Figure \ref{f4}a, which is for $\epsilon^\ast=0$ and therefore
corresponds to the phase point at $\lambda = 0.3$ in Figure \ref{f2}, we find
that $C_V^\ast$ exhibits two peaks. The smaller peak at low temperature
($T_{N_BN_U}^\ast=0.40$) is for a $N_{B}$ $\leftrightarrow$ $N_U$ transition and
the sharper peak at higher temperature ($T_{N_UI}^\ast=1.10$) is for the
$N_{U}$ $\leftrightarrow$ $I$ transition. 
In Figure \ref{f4}b ($\epsilon^\ast=0.8$) the two peaks of $C_V^\ast$
become closer due to the presence of dipolar 
interaction ($T_{N_BN_U}^\ast=0.85$, $T_{N_UI}^\ast=1.175$).
 In Figure \ref{f4}c ($\epsilon^\ast=1.6$) a single sharp peak
in $C_V^\ast$ is obtained which has a greater height and also occurs
at a higher temperature ($T_{N_BI}^\ast = 1.30$) indicating the formation of
the Landau line. 
In Figure \ref{f5}a we find a single broad hump in the specific heat plot
 for $\lambda=0.40825$ and $\epsilon^\ast=0$ which is similar
to the earlier findings \cite{bis}.
Figures \ref{f5}b and \ref{f5}c exhibit the influence of the dipolar strengths
on $C_V^\ast$ vs $T^\ast$ curves at the Landau point. We see that with the
increase in dipolar strength the peak height increases and the transition
temperature $T_{N_BI}^\ast$ shifts towards higher value.

\begin{figure}
\begin{center}
\resizebox*{100mm}{!}{\includegraphics{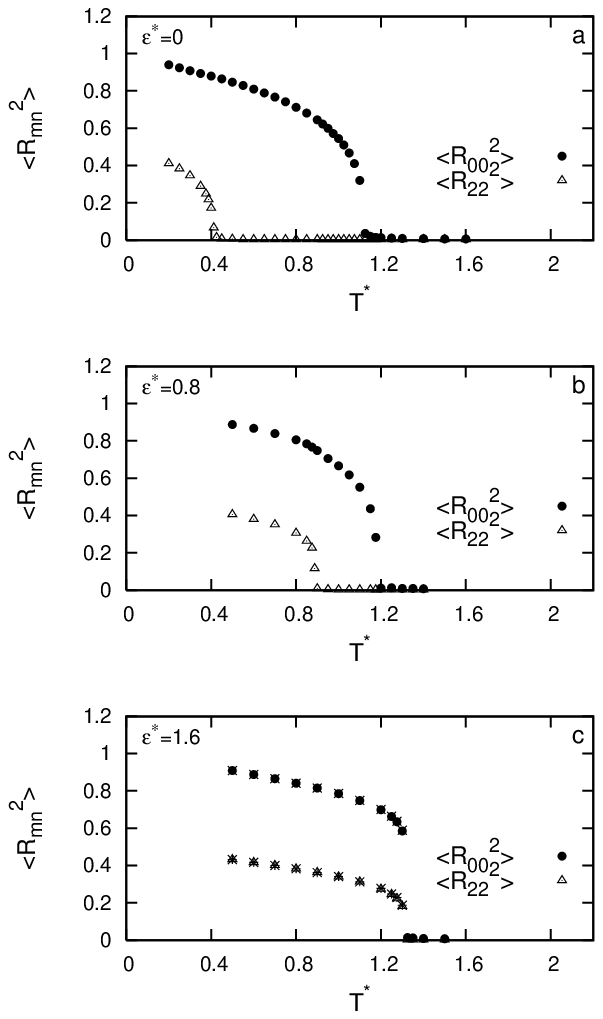}}
\caption{The average second rank (uniaxial and biaxial) nematic order
parameters $R_{mn}^2$ vs dimensionless temperature
$T^\ast$ for the molecular biaxiality parameter $\lambda=0.3$ and
for three different dipolar strengths $\epsilon^\ast= 0,0.8,1.6$.}\label{f6}
\end{center}
\end{figure}

\begin{figure}
\begin{center}
\resizebox*{100mm}{!}{\includegraphics{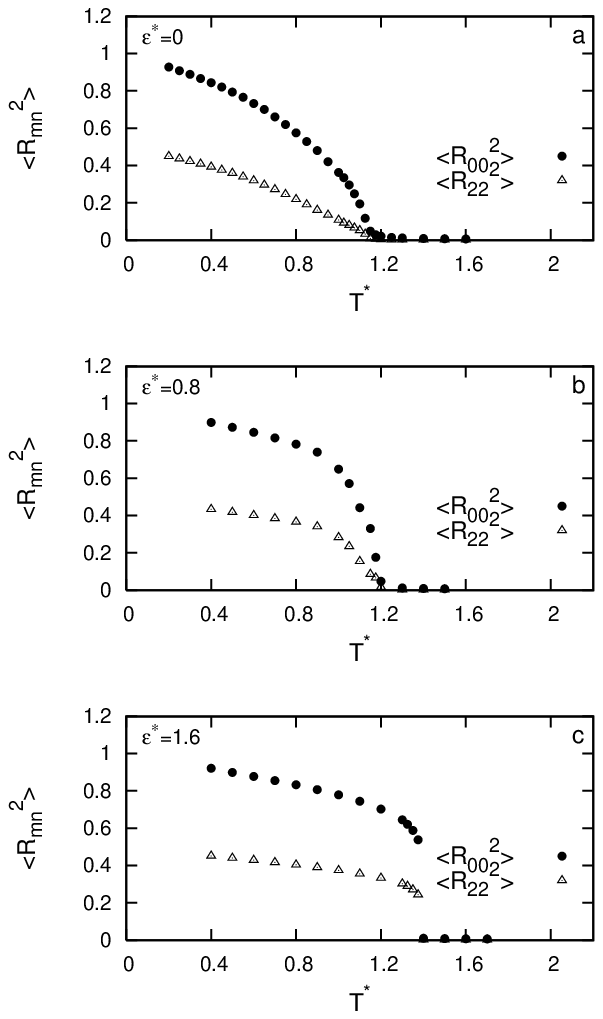}}
\caption{The average second rank (uniaxial and biaxial) nematic order 
parameters $R_{mn}^2$ vs dimensionless temperature  $T^\ast$ for
the molecular biaxiality parameter $\lambda=0.40825$ and for three
different dipolar strengths $\epsilon^\ast= 0,0.8,1.6$.}\label{f7}
\end{center}
\end{figure}

The temperature dependence of the second rank order parameters 
$\langle R_{00}^2\rangle$ and $\langle R_{22}^2\rangle$ 
for $\epsilon^\ast=0, 0.8$ and $1.6$ and for two different values of the 
biaxiality factor $\lambda =0.3$ and $0.40825$ are shown in Figure \ref{f6} 
and \ref{f7} respectively.
In Figure \ref{f6}a, which is for $\epsilon^\ast=0$ 
we find that the uniaxial and biaxial order parameters 
$\langle R_{00}^2\rangle$ 
and $\langle R_{22}^2\rangle$ represent order-disorder transitions at two
well separated temperatures.
The biaxial order parameter $\langle R_{22}^2\rangle$ decays 
continuously almost from its maximum value of $0.5$ to $0$ at the 
$N_{B}$ $\leftrightarrow$ $N_U$ transition ($T_{N_BN_U}^\ast=0.40$) whereas 
the uniaxial order parameter $\langle R_{00}^2\rangle$ decreases 
with the increase in $T^\ast$ and jumps 
to zero at the $N_{U}$ $\leftrightarrow$ $I$ transition ($T_{N_UI}^\ast=1.10$) 
which is an well known weakly first order transition. 
In Figure \ref{f6}b ($\epsilon^\ast=0.8$)  
both the $N_{B}$ $\leftrightarrow$ $N_U$ and $N_{U}$ $\leftrightarrow$ $I$
transitions get shifted towards higher $T^\ast$ and at the same time become
closer due to the presence of dipolar interaction ($T_{N_BN_U}^\ast=0.85$,
$T_{N_UI}^\ast=1.175$).  
In Figure \ref{f6}c ($\epsilon^\ast=1.6$)
both the transitions transform into a single 
direct $N_{B}$ $\leftrightarrow$ $I$ transition  
at a higher temperature ($T_{N_BI}^\ast = 1.30$).
The behaviour of $\langle R_{00}^2\rangle$ and $\langle R_{22}^2\rangle$ 
with $T^\ast$ in Figure \ref{f7}a for $\epsilon^\ast=0$ and $\lambda = 0.40825$ 
qualitatively reveal the continuous nature of the direct 
$N_{B}$ $\leftrightarrow$ $I$ transition ($T_{N_BI}^\ast=1.10$).  
The increase of strength of the 
dipolar interaction to $\epsilon^\ast=1.6$ (Figure \ref{f7}c) 
introduces a distinct change in character of the direct transition. 
In these figures sharp jumps in the values of $\langle R_{00}^2\rangle$ and
$\langle R_{22}^2\rangle$ from a considerable finite value to 
nearly zero values  
are observed as $T^\ast$ is gradually increased ($T_{N_BI}^\ast=1.375$).  

We have also calculated the first rank polar order parameter
$\langle P_1^u \rangle=\langle \boldsymbol{u}_i\cdot \boldsymbol{l}\rangle$
which is a measure of the degree of alignment of the transverse molecular 
symmetry axis $\boldsymbol{u}$ relative to the $\boldsymbol{l}$ direction 
of the director frame. We observe that this 
remains almost zero (within statistical error) over the entire 
temperature range for different values of the molecular biaxiality 
parameters and for all dipole strengths. 
Therefore, in any stable phase the system does not develop 
any spontaneous polarization. The ground state of the simple cubic system in
presence of the dipolar interaction ($U_{dd}$) alone
is anti-ferroelectric and hence does not produce any net overall polarization.
This is in agreement with the prediction made by Bates \cite{bates2} 
for a system of real dipolar bent core molecules although with the 
Heisenberg form of dipolar interaction used by Bates is likely to yield
a polar phase at low temperature for strong dipole moments.
It may be pointed out that the dipolar interaction used in 
reference \cite{bates2} is obtained by averaging out the anisotropic 
dipolar interaction over all orientations of the intermolecular 
vector for the nearest neighbours in a simple cubic lattice
whereas our finding of the absence of a polar phase is obtained 
with the full form of the dipole-dipole interaction including its long 
range character.

\begin{figure}
\begin{center}
\resizebox*{100mm}{!}{\includegraphics{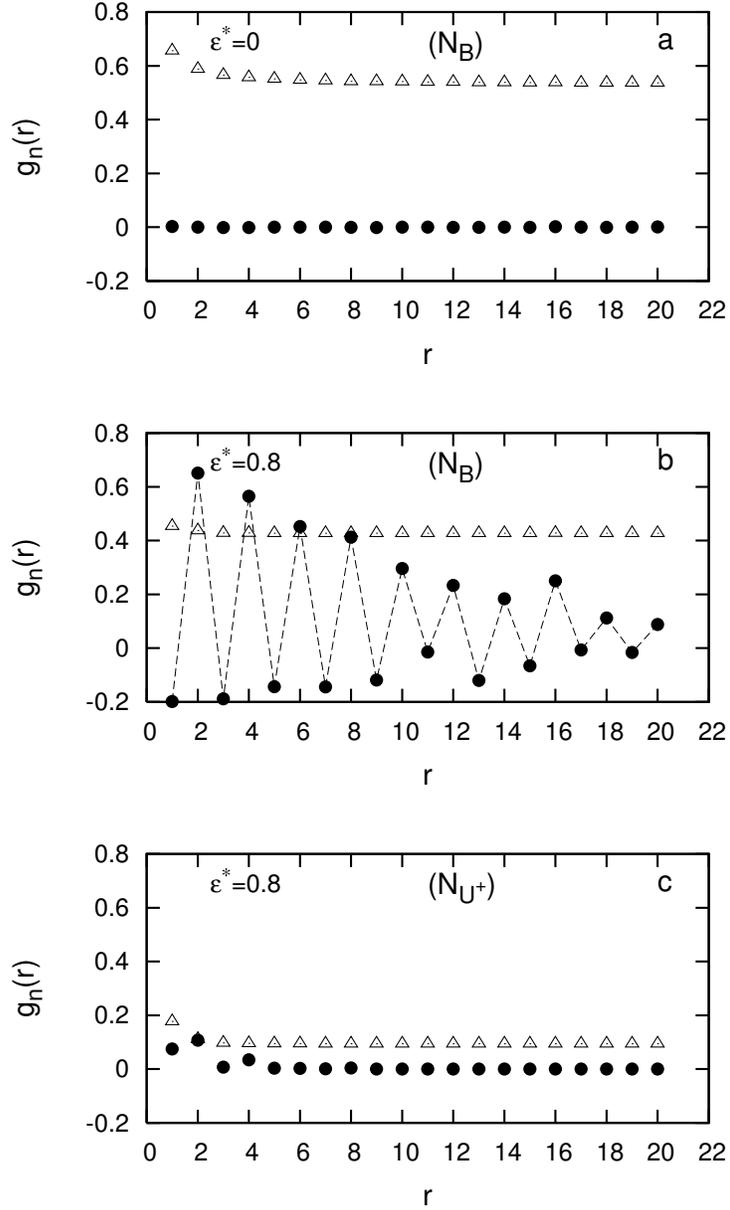}}
\caption{The first-rank and second-rank orientational correlation functions
  $g_1(r)$ (circles) and $g_2(r)$ (triangles) for the system of biaxial
molecules having biaxiality $\lambda=0.3$ with and without 
transverse dipoles at the temperatures ($a$)$T^\ast=0.3$ ($N_B$), 
($b$)$T^\ast=0.8$ ($N_B$) and ($c$)$T^\ast=1.0$ ($N_{U^+}$).
The dashed lines are used (in ($b$)) as a guide for the eye.}\label{f8}
\end{center}
\end{figure}

The orientational correlations between the transverse molecular
symmetry axes are represented by plotting the first-rank 
and the second-rank orientational correlation functions $g_1(r)$ and $g_2(r)$, 
defined in Section \ref{compu}, as a function of molecular separation 
$r$ in Figure \ref{f8} for the molecular biaxiality
$\lambda=0.3$. In case of pure dispersion interaction ($\epsilon^\ast=0$)
 the $N_B$ phase at lower temperature ($T^\ast=0.3$) has a long-range 
behaviour of $g_2(r)$ (Figure \ref{f8}a). The long-range limit of 
$g_2(r)$ ($0.537$) agrees with the square of the corresponding 
order parameter (which is $0.728$), according to the relation     
\begin{equation}\label{e18}
\lim_{r\to\infty}g_2(r) = (\langle \frac{3}{2}(\boldsymbol{u_i}\cdot \boldsymbol{l})^2-\frac{1}{2}\rangle)^2.
\end{equation}
We see that $g_1(r)$  is zero for all values of $r$ which shows, as can be 
anticipated, that the transverse molecular symmetry axis $\boldsymbol{u}$ 
has head tail symmetry in absence of transverse dipoles.

At higher temperature ($T^\ast=0.8$), the long range order of the 
anti-ferroelectric structure in the $N_B$ phase, induced by dipolar 
interaction ($\epsilon^\ast=0.8$), is observed from the pronounced
oscillations of $g_1(r)$ as shown in Figure \ref{f8}b.  
The positive value of $g_1(r)$ indicates a parallel arrangement 
of the pair of dipoles whereas its negative value shows their 
antiparallel arrangement. The oscillations of $g_1(r)$ is non-symmetric
about the zero axis because of the typical columnar anti-ferroelectric
structure of the simple cubic dipolar system. In a perfectly 
ordered state out of the six 
nearest-neighbours ($r=1$) of a dipole two dipoles are parallel while 
four are antiparallel and therefore $g_1(r)$ is $-1/3$. 
For the case of next nearest-neighbours ($r=2$)
all six dipoles are parallel and therefore $g_1(r)$ is $1$.
For this case the long range behaviour of $g_2(r)$ 
(Figure \ref{f8}b) is observed and its long range limit ($0.427$) is also found 
to agree with the square of the corresponding order-parameter 
($0.642$) according to Eq. (\ref{e18}). 
 
In Figure \ref{f8}c we see that in the $N_{U^+}$ phase ($T^\ast=1.0$) in spite 
of the presence of dipolar interaction ($\epsilon^\ast=0.8$) no 
anti-ferroelectric structure results. This is observed from the 
absence of long-range behaviour of $g_1(r)$. $g_1(r)$ remains zero for 
all values of $r$ except within the first two neighbours where a small 
oscillation occurs. Also from the variation of $g_2(r)$ it is found that 
no long-range correlation of the transverse molecular axis is present 
which confirms the absence of biaxial order.
Due to finite size effect a non-zero value ($\sim 0.1$) of $g_2(r)$
is observed at large $r$ which does not correspond to any true long range 
order for this case.
 
Since in our study the dipolar part of the total pair potential depends 
on the orientation of the intermolecular vector, we have found, 
as expected from a previous study \cite{hump}, 
that the directors are pinned along the lattice axes.
The interaction energy of a pair of dipoles depends on the angle
between the dipoles and the angles made by the dipoles 
with the vector joining them.
The lowest energy for an interacting pair of dipoles occurs 
when they are collinear and directed along the vector 
joining them. The next minimum occurs when they are   
antiparallel to each other but perpendicular to the inter-dipolar vector.
Therefore, for a simple cubic lattice system the ground state is a 
columnar anti-ferroelectric where 
the dipoles of a given row are parallel with one of the lattice axes
and nearest-neighbour columns are antiparallel.
Hence, due to the presence of anisotropy of dipolar interaction 
in orientation space the director frame 
coincides with the laboratory frame within a permutation of axes,
or in other words, the principal 
and the secondary directors are pinned along the lattice axes.       
In order to verify the above fact we have evaluated the averaged
quantities $\langle R_{00}^2\rangle_{lab}$ and 
$\langle R_{22}^2\rangle_{lab}$ which represent the uniaxial and biaxial 
order parameters with respect to the laboratory frame. 
The above quantities are calculated simply by taking the averages of 
$R_{00}^2$ and $R_{22}^2$ over 
the total number particles (i.e. the sample averages) for a given 
configuration and then by taking averages of these sample averages 
over large number of equilibrium configurations.  
The order parameters evaluated in the laboratory frame  
and their counterparts evaluated in the director frame differ by less 
than $2\%$.  
We have determined  $\langle R_{mn}^2\rangle_{lab}$ for 
$\lambda=0.3$ and $0.40825$ and dipolar strength $\epsilon^\ast=1.6$ for 
different values of the temperature in ordered state.
We have noticed that the directors remain pinned for all the 
temperatures and even close to the 
$N_{B}$ $\leftrightarrow$ $I$ transition they remain firmly pinned. 

\section{Conclusion}
The phase diagrams of biaxial molecules possessing $D_{2h}$ 
symmetry and with transverse dipoles, along their shortest 
dimensions, have been 
studied using Monte Carlo simulation. 
As already elaborated our results are in good qualitative agreement 
with those in references \cite{bates2} and \cite{grzy}.
We find that the phase 
diagram changes significantly due to the influence of strong (transverse)
dipolar interaction.
The most significant effect of the presence of  transverse dipoles 
is the splitting of the Landau point into a Landau line. 
Therefore, a range of geometrical structures of biaxial molecules 
are able to satisfy the most optimum condition of molecular biaxiality.    
We find that there is an optimal relative dipolar 
strength for $\mu\sim3.3$D for which 
the width of the Landau line becomes maximum and for this optimal value 
the dipolar interactions have almost an equal contribution of energy 
with that from the dispersion interactions at low temperature.
With further increase in dipolar strength the line begins to shorten 
and get shifted towards higher molecular biaxiality parameters.
As the strength of the dipole is increased, the direct 
$N_{B}$ $\leftrightarrow$ $I$ transition temperature increases  
thus leading to an increase in the stability of the thermotropic 
$N_B$ phase. The continuous nature of the direct 
$N_{B}$ $\leftrightarrow$ $I$ transition at the Landau point for the apolar 
biaxial system also appears to change because of the dipolar 
interaction. The sharp rise in the height of the specific heat peak
and the sudden jumps in the order parameters at the Landau point 
due to strong dipolar interactions tend to indicate qualitatively 
that the nature of the transition becomes first order.
A finite size scaling analysis is necessary to confirm this change 
in character of the phase transition. But considering the long-range 
nature of the dipolar interaction in simulations it becomes 
a very time consuming task and is therefore left out for future study.  
The structural properties in the uniaxial and biaxial phases are
investigated by evaluating first rank and second rank orientational
correlation functions. The dipole induced long range order of
the anti-ferroelectric structure in the $N_B$ phase is observed.
The lattice model that we have used is very simple in comparison 
to real thermotropic liquid crystalline systems 
because real mesogenic molecules have flexibility \cite{bates3} and also
possess both orientational and translational degrees of freedom. 

Finally we make a comment on the use of an anisotropic interaction 
(here the dipolar one) in a lattice model without taking an average 
over the orientations of the intermolecular vector. This is likely 
to result in orientationally ordered phases where the director becomes pinned
and the structure of the phases depend on the lattice symmetry. Our 
observation of the appearance of the dipole induced long-range 
anti-ferroelectric order may have resulted from this.
But we are not inclined to think that the biaxial nematic order 
we have observed in our work resulted from the use of a lattice model 
and the manner we have treated the dipolar interaction since 
the phases were identified by the relevant order parameters. A more 
elaborate study involving translational degrees of freedom of biaxial 
molecules with transverse dipole moment may perhaps be helpful.     

\section{Acknowledgment}
One of the authors (NG) acknowledges the award of a teacher fellowship
under Faculty Development Programme of the UGC, India.

\label{lastpage}
\end{document}